\documentstyle [proceedings]{crckapb}     
\def\lsim{{ {}^<_\sim}}

\def\dol{{d_{\rm ol}}}
\def\dls{{d_{\rm ls}}}
\def\dos{{d_{\rm os}}}

\begin{opening}

\title{\bf{Microlensing: Current Results and Future Prospects}}
\author{Andrew Gould}
\institute{Dept of Astronomy\\ 
Ohio State University, Columbus, OH 43210
\\ e-mail: gould@payne.mps.ohio-state.edu
\\{\rm Alfred P.\ Sloan Foundation Fellow}}
\end{opening}

\begin{document} 
\begin{abstract}
The initial results of microlensing surveys toward the Galactic bulge and
the LMC are puzzling.  Toward the LMC, the total mass in MACHOs is of order
half that required to explain the dark matter, but the estimated MACHO mass
($\sim 0.4\,M_\odot$) is most consistent with stars that should have been
seen.  Toward the bulge, the total mass is consistent with that given
by dynamical estimates of the bulge mass, but the observed timescales seem
to require that a large fraction of the bulge is made of brown dwarfs.  I
discuss possible experiments to resolve these puzzles. I also discuss
several applications of microlensing to non-dark-matter problems including
searching for planets and imaging the black hole at the center of a quasar
with $10^{-10}$ arcsec resolution.
\end{abstract}


\section{Introduction}

	Gravity bends light.  Any mass concentration therefore acts as a
gravitational lens.  (See Schneider, Ehlers, \& Falco 1992 for a review.)\ \ 
The magnification and distortion induced by this (or any)
lens can be described as a matrix $A_{i j}$, the transformation from the
source plane to the image plane.  By Liouville's (1837) theorem, surface 
brightness is conserved, so the magnification is given by the determinant
of this transformation, $A=|A_{i j}|.$

	Microlensing is, by definition, gravitational lensing where the
image structure is not resolved.  (See Paczy\'nski 1996; Roulet \& Mollerach
1997; and Gould 1996 for reviews.)\ \ 
The only observable is therefore the total
magnification from all the images.  The most important case is microlensing
of a point source by a point mass.  If the lens were perfectly aligned
with the observer-source line of sight, the source would be imaged into a
ring of angular radius, $\theta_e$, called the Einstein radius
\begin{equation}
\theta_e = \sqrt{4 G M\over D c^2};\qquad D\equiv {\dol\dos\over\dls},
\end{equation}
where $M$ is the mass of the lens, and $\dol$, $\dls$, and $\dos$ are the
distances between the observer, lens, and source.  If the alignment is not
perfect, there are two images, one inside and one outside the Einstein ring
and with magnifications
\begin{equation}
A_\pm = {A\pm 1\over 2};\qquad A(x) = {x^2+2\over x\sqrt{x^2+4}},
\end{equation}
where $x$ is the projected separation of the source and lens in units of
the Einstein ring.  

	The signature of microlensing is a specific form of time variability 
of the source flux.  If the lens moves with uniform transverse velocity
$\bf v$ relative to the observer-source line of sight, then the separation
$x$ is given by the Pythagorean theorem,
\begin{equation}
x(t) = \sqrt{{(t-t_0)^2\over t_e^2} + \beta^2},
\end{equation}
where $\beta$ is the impact parameter in units of $\theta_e$, $t_0$ is the
time of maximum magnification, and $t_e$ is the Einstein crossing time,
\begin{equation}
t_e = {\dol\theta_e\over v}. 
\end{equation}
Both the power and limitations of microlensing are summarized in these four
equations.  On the one hand, the microlensing light curve is described
by only three parameters, $t_0$, $\beta$, and $t_e$.  Thus microlensing can
be distinguished from other forms of stellar variability which are about 1000
times more common.  On the other hand, all of the information about the lens
is contained in $t_e$ which is itself a complicated combination of $M$,
$\dol$, and $v$.

	Here I will summarize some of the major results obtained from 
microlensing experiments to date and also discuss some future possibilities.

\section{Dark Matter and Dim Matter}

	Paczy\'nski (1986) proposed that one could search for Machos (lumps
of dark matter) in the halo of the Milky Way by monitoring stars in the Large
Magellanic Cloud (LMC) for microlensing events.  Astronomers reacted to this
the same way they usually do whenever someone comes up with a good new idea:
they said it was impossible.  The problem is that the probability that any
given star is microlensed at any given time is $<10^{-6}$, so that it would
be necessary to monitor millions of stars to get a few events.  This seemed
to be beyond the capabilities of individual observers.  However, particle
physicists were also interested in this problem because if the dark matter is
not made of Machos, it may be new fundamental particles.  The experimental
requirements are modest by particle physics standards.  Two groups
(MACHO, Alcock et al.\ 1997b; EROS, Ansari et al.\ 1996), each
composed of particle physicists and astronomers, initiated Macho searches
toward the LMC.  

	The initial results of these LMC searches are extremely puzzling.
MACHO finds a total of 8 candidate events from their first two years of data
with time scales $t_e$ ranging from about 2.5 weeks to 2.5 months.  While one
of these events is most likely due to a stellar lens in the disk of the Milky
Way (Gould, Bahcall, \& Flynn 1997), 
one is likely due to a binary-star lens in the LMC, and one is quite 
possibly a variable star (and thus not microlensing),  the remaining 5 events
are difficult to explain within the context of standard models of our Galaxy.
If they are due to dark objects in the halo, these objects seem to account
for $\sim 50\%$ of the dark matter.  The problem with this interpretation 
lies in the time scales.  From the microlensing equations of \S\ 1, it is
clear that $t_e\propto \sqrt{M}$ and so contains some information about mass,
but that it also depends of $\dol$ and $v$.  The distributions of these latter
quantities is fixed for any given halo model, so for a given mass, there
is a probability distribution of time scales.  The best estimate for the
mean Macho mass is $\sim 0.4\pm 0.2\,M_\odot$.  What could such objects be?
They cannot be made of primordial material (H and He) because they would be
M stars and easily visible.  In fact star counts show that halo M stars are
about 100 times less numerous than is required to solve the dark matter 
problem.  That is why there is a dark matter problem.  The estimated Macho 
mass is consistent
with white dwarf (WD) masses, but several arguments make this solution seem 
implausible.  Star counts place a 95\% confidence upper limit on the density
of WDs that is close to the
density required to explain the Macho events (Flynn, Gould, \& Bahcall 1996).  
Moreover, WD progenitors would
pollute the halo with metals and would `light up' distant galaxies with 
red-giant light if this were indeed the universal explanation of dark matter
(and not just in our Galaxy).  It is a mark of the paucity of other plausible
ideas to explain the observed microlensing events that the WD scenario is
taken seriously at all.

	An interesting set of alternative ideas is that the majority of the
events do not arise from the halo, but rather from stellar objects in the
LMC itself (Sahu 1994) or a thick disk in the Milky Way (Gould 1994a).
Unfortunately, dynamical constraints appear to limit the contributions of both
structures to levels 
that are well below the observed lensing rate (Gould 1995b).  In brief, there
are no sensible ideas to account for the observed lensing toward the LMC.

	At first sight, the situation appears quite different toward the 
Galactic bulge, where four groups have carried out lensing observations
(OGLE, Udalski et al.\ 1994; MACHO, Alcock et al.\ 1997a; DUO, Alard 1996;
EROS).  Since the line of sight passes through the disk and the bulge itself,
many events are expected from ordinary stars.  Many are seen and the time
scales are broadly consistent with stellar masses.  Zhao, Spergel, \& Rich
(1995) showed that the time scale distribution could be explained if the
dynamically measured mass of the bulge ($\sim 2\times 10^{10}\,M_\odot$)
were distributed in a Salpeter (power-law) mass function between 0.6 and
0.08$\,M_\odot$, the latter value being the hydrogen-burning limit.
Unfortunately, this simple picture does not hold up under closer examination.
 First, all of the bulge mass cannot be in objects $M<0.6\,M_\odot$, since the
bulge luminosity function (LF) has been measured for $M>0.5\,M_\odot$.  These
account for about half of the bulge mass but very few of the observed lensing
events (Han 1997).  If the bulge LF (Light, Baum, \& Holtzman 1997)
is extended according to the locally
measured disk LF (Gould, Bahcall, \& Flynn 1996), then about 2/3 of the bulge 
mass is accounted for, but hardly any of the short ($t_e\sim 10\,$day) 
lensing events.  Only if the last 1/3 of the bulge
mass is assumed to be in brown 
dwarfs can the bulge lensing observations be explained (Han 1997).

	In brief, both the LMC and bulge lensing observations are difficult 
to explain, but for opposite reasons.  The LMC events seem to require lenses
that are so massive ($\sim 0.4\,M_\odot$) that they should shine and be
noticed.  The bulge events seem to require a new population of substellar
objects not previously detected.

\section{Resolving the Macho Mysteries}

	Clearly, the best way to figure out what these objects are is to
determine their individual masses, velocities, and distances.  To date this
has not been possible because the only information available is the time scale,
$t_e$, which is a complicated combination all three: $t_e=t_e(M,\dol,v)$.
  Two additional pieces
of information are needed to fully break this degeneracy.  One parameter that
one might hope to measure is the size of the Einstein ring projected onto
the source plane.  Another is the Einstein ring projected onto the plane of
the observer.  These are respectively,
\begin{equation}
\hat r_e \equiv \dos\theta_e; \qquad \tilde r_e\equiv D\theta_e.
\end{equation}
(Note that, since $\dos$ is generally known reasonably well, determining
$\hat r_e$ is equivalent to determining $\theta_e$.)\ \
In either case, there must be some standard ruler in the source plane or in the
observer plane and there must be some effect that depends on the size of the
Einstein ring relative to that ruler.  If both parameters were measured, then
one could determine $M$, $\dol$, and $v$.  For example,
\begin{equation}
M = {c^2\over 4 G}\tilde r_e\theta_e.
\end{equation}
Even if only one of these two quantities were measured for a large sample
of events, the character of the events would be substantially clarified.
 For example, if $\tilde r_e$ were measured, then one would also know
the ``projected speed'',
\begin{equation}
\tilde v = {\tilde r_e\over t_e} = {\dos\over\dls}v.
\end{equation}
This quantity is $\sim 50\,\rm km\,s^{-1}$ for disk lenses,
$\sim 300\,\rm km\,s^{-1}$ for halo lenses, and
$\sim 2000\,\rm km\,s^{-1}$ for LMC lenses.  Hence these populations could
be easily separated.

	Where is one to find these standard rulers?  By far, the best
plan would be to create such a ruler in the plane of the observer by
launching a parallax satellite into solar orbit (Refsdal 1966; Gould 1994b,
1995a; Boutreux \& Gould 1996; Gaudi \& Gould 1997a).  Since $\tilde r_e\sim
{\cal O}$(AU), there is a significant fractional vector displacement in the 
Einstein ring of the event as seen from the satellite relative to the Earth,
${\bf \Delta x} = {\bf d}_{\rm sat}/\tilde r_e$, where ${\bf d}_{\rm sat}$
is the position of the satellite relative to the Earth.  Hence, 
by measuring the
difference in impact parameters $\Delta \beta=\beta'-\beta$ and 
difference in times of maximum $\delta t=t_0'-t_0$ between the event as seen
from the satellite and from the Earth, one can determine
${\bf \Delta x}= (\Delta t/t_e,\Delta\beta)$ and so $\tilde r_e$.
Another method to measure the same quantity is to use the Earth's orbit as 
baseline (Gould 1992b).  Unfortunately, most events end before the Earth
has moved far enough to generate a significant effect. Nevertheless, 
$\tilde r_e$ has been measured for one event using this method (Alcock 1997a).

	The most ubiquitous standard ruler in the source plane is the source
itself whose angular radius $\theta_*$ is known from its color, magnitude, and 
Stefan's Law.  If the lens transits the source (at say, $x=x_*$), 
the light curve 
will deviate from its standard form and one can therefore measure 
$\theta_e=\theta_*/x_*$ (Gould 1994a;
Nemiroff \& Wickramasinghe 1994; Witt \& Mao 1994).  Unfortunately, the 
fraction of events for which this is possible is only 
$\sim \theta_*/\theta_e$, i.e., $<5\%$ for the bulge and $<1\%$ for the LMC.
 For bulge events, there are a variety of other methods to measure $\theta_e$,
notably optical/infrared photometry (Gould \& Welch 1996) and infrared
interferometry (Gould 1996).  For the LMC, however, there are only two methods
known that could plausibly provide information about $\theta_e$.  First, if
the source happens to be a binary, the separation between the stars can be 
used as a standard ruler which
is enormously larger than the physical extent of the individual stars
(Han \& Gould 1997).  Second, lensing of rapidly rotating sources 
(like A stars) creates an apparent line shift because the redshifted side
of the star is magnified by a different amount than the blueshifted side.  This
allows one to measure $\theta_*/\theta_e$ even if the impact parameter is
many source radii (Maoz \& Gould 1994).

	In general, the various techniques discussed in this section require
significant investments in observational effort and/or money.  However, the
methods are practical and well within present capabilities.  It is possible
to figure out what the lenses are if we make the effort.

\section{Pixel Lensing of M87}

	For the remainder of the presentation, I will focus on what the
future of microlensing may look like.  This is highly speculative, but for
such a new and rapidly developing field, rampant speculation is quite in order
... and may even prove productive.  I begin by discussing possible microlensing
of M87, the central galaxy in the Virgo cluster.  This raises two immediate
questions: how is it possible to observe microlensing events in a galaxy whose
stars are completely unresolved, and why bother to observe them anyway?

	I will not spend much time discussing how microlensing of unresolved
stars can be observed.
Most people even in the microlensing business considered that it was impossible
when it was first proposed by Crotts (1992) and Baillon et al.\ (1993).  But
both groups have now demonstrated its feasibility in observations toward M31
(Tomaney \& Crotts 1996; Ansari et al.\ 1997).  Crotts will describe this 
success immediately following my presentation.

	Rather let me focus on why M87 is an especially interesting target.
The microlensing results from MACHO and EROS indicate that perhaps half of
the dark matter in the Milky Way halo is in Machos.  That is, the total mass
of Machos is equal to or perhaps twice as large as the mass of all the stars in
the known components of the Galaxy, the bulge and the disk.  Hence, one might
suppose that as the Galaxy was forming and was still roughly spherical, 
half of the available gas was processed into Machos.  The remaining gas 
collapsed into a proto-disk and proto-bulge which went on to form the
visible Galaxy we know and love.  Imagine then a Milky-Way like galaxy forming
on the outskirts of the Virgo cluster.  Like the real Milky Way, it would
process half of its gas into Machos with the remaining gas beginning to 
collapse into a proto-bulge and proto-disk.  But before these collapsed
gas clouds could form many stars, the galaxy would fall through the center
of the cluster and would 
be stripped of its gas by the hot intracluster medium.  
The galaxy would become a dark Macho galaxy.  The total mass in Machos would
be about equal to the mass in gas, i.e., 20\% or so of the total mass of the
cluster.  The Macho galaxy
might remain intact or dissolve, but in either event its 
Machos would give rise to microlensing events of M87.  Detection of these
events from the ground is probably not possible, but 10 continuous
days of observations by the Hubble Space Telescope (HST) would yield $\sim 30$
events and so test this scenario directly (Gould 1995e).

	There are many other applications of microlensing outside the Local
Group, such as probing the star formation history of the universe (Gould 1995d)
and measuring the transverse velocities of distant galaxies (Gould 1995c).
However, time is short so I move on to an application closer to home.

\section{Planet Detection}

	Microlensing can be used to detect anything that is dark.  One 
interesting possibility is planets (Mao \& Paczy\'nski 1991; Gould \& Loeb
1992).  Suppose that a star is being microlensed by another star.  Such events
happen frequently toward the bulge.  The light from the source star comes to
us along two paths, one on either side of the lensing star.  If the lensing
star has a planet, and one of the light trajectories happens to come near
that planet, then the planet will further deflect the light causing a deviation
of the light curve from the standard form discussed in \S\ 1.  The deviation
will be shorter than the event as a whole by a factor $\sqrt{m_p/M}$ where
$m_p$ is the mass of the planet.  That is, the deviation will most likely last
less than a day for a Jupiter-mass planet or smaller.  It might therefore be
missed by the ordinary microlensing search observations since these are
typically carried out only once per day.  However, the size of the deviation
will typically be large, so that if the deviation is observed repeatedly, there
will be no question that a planet has been detected.  Hence, one should attempt
to organize round-the-clock (i.e., round-the-world) observations once every
few hours to catch such events.  Two groups have begun such follow-up 
observations using observatories in Chile, South Africa, Israel, Australia,
and New Zealand (Albrow et al.\ 1996; Pratt et al.\ 1996).  Substantial
improvements in these observations are expected when two optical/infrared
cameras are placed on near-dedicated telescopes to join this follow-up
program (D.\ DePoy 1997, private communication).  The theoretical problems
associated with the analysis of planetary light curves initially seemed rather
daunting because the planetary Einstein ring is generally of the same size as
the source star.  However, substantial progress is now also being made on
this front as well (Bennett \& Rhie 1996; Gaucherel \& Gould 1997; Gaudi \& 
Gould 1997b).

\section{Femtolens Imaging of Quasar Black Holes}

	Microlensing is developing with incredible speed.  One indication of 
this is that while most of the ideas discussed in the previous sections
were considered ``crackpot'' (or more politely, 
``too advanced for their time'') when they were first proposed, many led
almost immediately to new observational programs and the detection of new
effects.  Paczy\'nski's (1986) original microlensing proposal is the most 
famous example of this, but there are many others.  I already discussed the 
rapid implementation of the Crotts (1992) and Baillon et al.\ (1993) idea for 
pixel lensing.  Finite source effects and ground-based parallax
were both observed within 2 years of first being predicted.  The proposal
to search for planets was taken up by two world-wide collaborations within
3 years.  The idea for a parallax satellite became a NASA proposal
and pixel lensing of M87 became an HST proposal, both within 1 year (although
neither is yet successful).  If the most outrageous ideas that theorists
can invent come to pass within a couple of years, then certainly we are not
being imaginative enough!  Here I present an attempt to overcome this 
shortcoming:  femtolens interferometry of quasar black holes.

	To explain femtolens interferometry, I must first describe simple
femtolensing.  Recall from \S\ 1 that for a simple point-mass lens, there
are two images.  When I calculated the total magnification of the lens, I
simply added the two magnifications together, $A=A_+ + A_-$.  However, if the
point source is truly a point, then the two images will arrive separated
by a time delay $\Delta t$.  To a good approximation
\begin{equation}
\Delta t(x)\simeq {8 G M\over c^3} x.
\end{equation}
Hence, for light at wavelength $\lambda$, there will also be a phase delay
$\phi = c\Delta t/\lambda$, and the true magnification will be
\begin{equation}
A = {\cal A}_+\cos^2{\phi\over 2} +{\cal A}_-\sin^2{\phi\over 2};\qquad
{\cal A}_\pm = (A_+^{1/2}\pm A_-^{1/2})^2=
\biggl(1 + {4\over x^2}\biggr)^{\pm 1/2}.
\end{equation}
Normally interference is not important because real sources are so big that
interference effects at different points on the source have different phases
which cancel one another out.  However, if $\gamma$-ray bursts come from
cosmological distances, and if they were lensed by asteroid-mass objects,
their spectra would show oscillations with peak-to-trough variations of
${\cal A}_+/{\cal A}_-=(1 +4/x^2)$, which would easily be noticed 
(Gould 1992a).  Thus, $\gamma$-ray 
bursts could be used to probe for or put limits on such objects.

	Femtolens interference can be extended to femtolens interferometry,
but some additional investment is required (Gould \& Gaudi 1997).  First one
must find a nearby $(\lsim 30\,$pc) dwarf star that is perfectly aligned with
a distant quasar.  The star is to serve as the ``primary lens'' of a giant
telescope to image the quasar.  Unfortunately, even 
if such an alignment happened
to occur, the transverse motion of the dwarf ($\sim 40\,\rm km\, s^{-1}$)
would wreck the telescope as soon as it was set up.  So it will be necessary
to use a satellite to bring the ``secondary optics'' of the telescope into
alignment with the dwarf-quasar line of sight ... and keep it there.  There
should be such a point of alignment within $\sim 45\,$AU of the Sun.  If the
dwarf star were isolated, the quasar would be imaged into two images.  However,
most dwarfs have binary companions.  Such a companion is just what is needed
to create a femtolens imaging telescope.  It creates an ``astigmatism'' in the
lens called a ``caustic''.  If the quasar lies inside the caustic, then there
are 5 images (instead of two for a point lens).  One of these images is close
to companion and will be ignored.  If the quasar lies close to a cusp of the
caustic, then three of the remaining images will be very highly magnified
and lie on one side of the dwarf, while the fourth image will be only 
moderately magnified and lie on the opposite side of the dwarf.  It will be
ignored.  The typical magnifications of the three images are $\sim 10^6$ in
one direction, but there is an actual demagnification by a factor of 2 in the
other direction.  That is, each image will be highly elongated: it can be
resolved in one direction, but not the other.  For example, if a quasar black
hole has a mass $M\sim 10^8\,M_\odot$, then its Schwarzschild radius is 
$\sim 1\,$AU.  At a cosmological distance it therefore subtends $10^{-9}$
arcsec.  Its image will then be 1 mas $\times 10^{-6}\,$mas.  The first
dimension is easily resolved with a space-based telescope.  The second is not.
The point of femtolens interferometry is to resolve the second dimension.

	If the image of the quasar is resolved in one dimension, then light
from different portions of the image can be brought together and analyzed in
a spectrograph.  Each portion will contain light from a one-dimensional 
strip through the quasar.  These strips generally intersect one another only
in a limited region.  If the two portions are brought together, then only
the light from this limited region suffers interference.  Actually, each
such region contains subregions with different relative time delays between
the two image portions.  The interference pattern is the Fourier transform
of the this time-delay structure.  It can reveal structure as small as
$\sim 1/10$ AU, which is 
the separation in the source plane at which the relative time
delay differs by one $\lambda/c$ where $\lambda$ is the typical wavelength of 
optical light.

	Of course, there are a few engineering problems associated with this
idea.  It is easy to get a satellite to 45 AU, but this one must be given
an additional boost of $\sim 40\,\rm km\, s^{-1}$ once it gets there.  The
mirror system must extend about 350 m in a one-dimensional array 
in order to re-image the gravitationally lensed quasar images.  This is not
out of line with other plans for space-based interferometers.  However, in
this case, the mirror system must be accelerated by $\sim 20\,\rm cm\, s^{-2}$
about once every 10 hours in order to counter the Sun's gravity, and the
mirror system must restablize after each such jolt.  However, microlensing
has met previous challenges and I am confident it will meet these as well.

\section{Conclusion}

	More good microlensing ideas are needed.

\bigskip

This work was supported in part by NSF grant AST 9420746.

\bigskip
\centerline{\bf References}
\smallskip

\begin{enumerate}
\item{} Alard, C.\ 1996, in IAU Symp.\ 173 ed.\ C.\ S.\ Kochanek \&
J.\ N.\ Hewitt) (Dordrecht: Kluwer), 215
\item{} Albrow, M., et al.\ 1996, in IAU Symp.\ 173 ed.\ 
C.\ S.\ Kochanek \& J.\ N.\ Hewitt) (Dordrecht: Kluwer), 227
\item{} Alcock, C., et al.\ 1997a, ApJ, in press
\item{} Alcock, C., et al.\ 1997b, ApJ, in press
\item{} Ansari, R., et al.\ 1996, A\&A, 314, 94
\item{} Ansari, R., et al.\ 1997, A\&A, in press
\item{} Baillon, P., Bouquet, A., Giraud-H\'eraud, Y., \& Kaplan, J.\ 1993
A\&A, 277, 1
\item{} Bennett, D.\ P., \& Rhie, S.\ H.\ 1996, ApJ, 472, 660
\item{} Boutreux, T., \& Gould, A.\ 1996, ApJ, 462, 705
\item{} Crotts, A.\ P.\ S.\ 1992, ApJ, 399, L4
\item{} Flynn, C., Gould, A., \& Bahcall, J.\ N.\ 1996, ApJ, 466, L55
\item{} Gaucherel, C., \& Gould, A.\ 1997, ApJ, 477, 580
\item{} Gaudi, B.\ S., \& Gould, A.\ 1997a, ApJ, 477, 152
\item{} Gaudi, B.\ S., \& Gould, A.\ 1997b, ApJ, submitted
\item{} Gould, A.\ 1992a, ApJ, 386, L5
\item{} Gould, A.\ 1992b, ApJ, 392, 442
\item{} Gould, A.\ 1994, ApJ, 421, L71
\item{} Gould, A.\ 1994, ApJ, 421, L75
\item{} Gould, A.\ 1995a, ApJ, 441, L21
\item{} Gould, A.\ 1995b, ApJ, 441, 77
\item{} Gould, A.\ 1995c, ApJ, 444, 556
\item{} Gould, A.\ 1995d, ApJ, 455, 37
\item{} Gould, A.\ 1995e, ApJ, 455, 44
\item{} Gould, A.\ 1996, PASP, 108, 465
\item{} Gould, A., Bahcall, J.\ N., \& Flynn, C.\ 1996, ApJ, 465, 759
\item{} Gould, A., Bahcall, J.\ N., \& Flynn, C.\ 1997, ApJ, 482, 000
\item{} Gould, A., \& Gaudi, B.\ S.\ 1997, ApJ, submitted
\item{} Gould, A., \& Loeb, A.\ 1992, ApJ, 396, 104
\item{} Gould, A., \& Welch, D.\ L.\ 1996, ApJ, 464, 212
\item{} Han, C.\ 1997, ApJ, in press
\item{} Han, C.\ \& Gould, A.\ 1997, ApJ, 480, 000
\item{} Light, R.\ M., Baum, W.\ A., \& Holtzman, J.\ A.\ 1997, 
in preparation
\item{} Liouville, J.\ 1837, Journal de Math\'ematiques Pures et 
Appliqu\'ees, 2, 16
\item{} Mao, S., \& Paczy\'nski, B.\ 1991, ApJ, 388, L45
\item{} Maoz, D., \& Gould, A.\ 1994, ApJ, 425, L67
\item{} Nemiroff, R.\ J.\ \& Wickramasinghe, W.\ A.\ D.\ T.\ 1994, ApJ, 
424, L21
\item{} Paczy\'nski, B.\ 1986, ApJ, 304, 1
\item{} Paczy\'nski, B.\ 1996, ARAA, 34, 419
\item{} Pratt, M., et al.\ 1996, in IAU Symp.\ 173 ed.\ 
C.\ S.\ Kochanek \& J.\ N.\ Hewitt) (Dordrecht: Kluwer), 221
\item{} Refsdal, S.\ 1966, MNRAS, 134, 315
\item{} Roulet, E.\ \& Mollerach, S.\ 1997, Physics Reports, in press
\item{} Sahu, K.\ C.\ 1994, Nature, 370, 275
\item{} Schneider, P., Ehlers, J., \& Falco, E.\ E.\ 1992, Gravitational 
Lenses (Berlin: Springer-Verlag)
\item{} Tomaney, A., \& Crotts, A.\ P.\ S.\ 1996, AJ, 112, 2872
\item{} Udalski, A., et al.\
  1994, Acta Astron, 44, 165
\item{} Witt, H., \& Mao, S.\ 1994, ApJ, 430, 505
\item{} Zhao, H.\ S., Spergel, D.\ N., \& Rich, R.\ M.\ 1995, ApJ, 440, L13
\end{enumerate}
\end{document}